# Circularly polarized electroluminescence from a single-crystal organic microcavity light-emitting diode based on photonic spin-orbit interactions

Jichao Jia,[1] Xue Cao,[1] Xuekai Ma,[2] Jianbo De,[3] Jiannian Yao,[3] Stefan Schumacher,[2,4] Qing Liao,[1,*] Hongbing Fu[1,*]

[1]Beijing Key Laboratory for Optical Materials and Photonic Devices, Department of Chemistry, Capital Normal University, Beijing 100048, People's Republic of China

[2]Department of Physics and Center for Optoelectronics and Photonics Paderborn (CeOPP), Universität Paderborn, Warburger Strasse 100, 33098 Paderborn, Germany

[3]Institute of Molecule Plus, Tianjin University, and Collaborative Innovation Center of Chemical Science and Engineering (Tianjin), Tianjin 300072, P. R. China

[4]Wyant College of Optical Sciences, University of Arizona, Tucson, Arizona 85721, United States

**Abstract**

Circularly polarized (CP) electroluminescence from organic light-emitting diodes (OLEDs) has aroused considerable attention for their potential in future display and photonic technologies. The development of CP-OLEDs relies largely on chiral-emitters, which not only remain rare owing to difficulties in design and synthesis but also limit the performance of electroluminescence. When the polarization (pseudospin) degrees of freedom of a photon interact with its orbital angular momentum, photonic spin-orbit interaction (SOI) emerges such as Rashba-Dresselhaus (RD) effect. Here, we demonstrate a chiral-emitter-free microcavity CP-OLED with a high dissymmetry factor ($g_{EL}$) and high luminance by embedding a thin two-dimensional organic single crystal (2D-OSC) between two silver layers which serve as two metallic mirrors forming a microcavity and meanwhile also as two electrodes in an OLED architecture. In the presence of the RD effect, the SOIs in the birefringent 2D-OSC microcavity result in a controllable spin-splitting with CP dispersions. Thanks to the high emission efficiency and high carrier mobility of the OSC, chiral-emitter-free CP-OLEDs have been demonstrated exhibiting a high $g_{EL}$ of 1.1 and a maximum luminance of about 60000 cd/m$^2$, which places our device among the best performing CP-OLEDs. This strategy opens a new avenue for practical applications towards on-chip microcavity CP-OLEDs.

## Introduction

Circularly polarized (CP) light, featuring optical rotatory power and rich angle-independent properties, has attracted increasing attention for a variety of potential applications, such as encrypted information storage[1,2], three-dimensional (3D) displays[3-7], remote sensing[8,9], and chiroptical switches[10-13]. In the case of organic light-emitting diode (OLED) based 3D-display technology, incorporation of extra polarizers is required to produce CP light from unpolarized electroluminescence (EL), leading to large power loss and poor contrast ratio[14,15]. Therefore, the active generation of CP light directly from OLEDs, i.e. CP-OLEDs, is more practical, due to their simple device architectures and energy saving without the need for extra polarizers[5,6,16,17]. So far, the dominant efforts for CP-OLEDs focus on the development of chiral emitters as the active layer for CP EL, including chiral molecules and metal-organic complexes[3,18-20], as well as achiral conjugated polymers with chiral sidechains and chiral dopants[4,21-24]. Nevertheless, high-performance CP-OLEDs remain a great challenge hindered by three stumbling blocks: (i) The CP EL materials are rare, limited by difficult molecular design and synthesis[25]; (ii) The key parameter, dissymmetry factor ($g_{EL}$) of the CP-OLEDs, is still relatively low, in the range of $10^{-3} \sim 10^{-1}$, except for a few examples which exhibit high $g_{EL}$ up to 1.0 by using chiral transition metal complexes and polymers[4,14,26]; (iii) The CP-OLEDs still show lower device performance than conventional OLEDs, for example, the relatively low luminance, inevitably hindering their practical applications[13]. Therefore, the development of CP-OLEDs with high $g_{EL}$, high luminance and of those that are chiral-emitter free is a key issue to be addressed.

The polarization degree of freedom is intrinsic to the nature of light and it can interact with the light field's orbital angular momentum. This is called photonic spin-orbit interaction (SOI) and is reminiscent of SOI in electronic systems with the photon's pseudospin mimicking the electron's spin. Photonic SOI has been widely investigated in inorganic systems, such as graphene, transition-metal dichalcogenide, and metasurface materials[27-29], and brings about abundant new applications in optoelectronics ranging from classical information processing to the quantum optical regime[30-32]. The realization of photonic SOI requires the breaking of inversion symmetries in solid-state systems. Contrary to inorganic materials, organic molecular assemblies, especially organic single crystals (OSCs), have highly ordered and anisotropic molecular packing arrangement and therefore anisotropic refractive index and show birefringence. Recently, the tunability of both energy and polarization of the confined photonic modes have been reported in liquid-crystal-filled[33] and OSC-filled[34] birefringent organic cavities. In particular, when two photonic modes with orthogonal linear-polarization and opposite parity are close to resonance, Rashba-Dresselhaus (RD) SOI emerges with the characteristic feature of left- and right-handed CP dispersions[33,34]. Regrettably, the CP-splitting phenomenon due to RD-SOI is demonstrated in most cases in passive reflection mode. We anticipate that combining a birefringent OSC cavity with OLED architecture might lead to the realization of artificial RD SOI, producing CP EL without the need of chiral emitters.

In the present work, we demonstrate a chiral-emitter free CP-OLED with high $g_{EL}$

and high luminance by embedding a thin two-dimensional OSC (2D-OSC) of 1,4-bis((E)-2,4-dimethylstyryl)-2,5-dimethylbenzene (6M-DSB) sandwiched between two silver layers which serve as two mirrors forming a planar microcavity and meanwhile as two electrodes in OLED configuration. Introducing birefringent 6M-DSB 2D-OSC into a planar microcavity modulates the polarization of the confined photonic modes through artificial RD SOI. Thanks to high emission efficiency and high carrier mobility of 6M-DSB OSC[35-41], chiral-emitter free CP-OLEDs are demonstrated with a high $g_{EL}$ of 1.1 and a maximum luminance of about 60000 cd/m$^2$, which are among the best performances of CP-OLEDs. This unique top-emitting microcavity OLED architecture with active CP EL based on RD SOI demonstrates a promising strategy for future display and communication applications.

**Result and discussion**

A schematic of our device is shown in Fig. 1a. The microcavity CP-OLEDs were fabricated with a top-emitting device architecture: silicon wafer/Ag (200 nm)/6M-DSB OSC (990 nm)/CsF (10 nm)/Ag (35 nm) (see details in the Supplementary), where the bottom Ag and the top CsF/Ag films played the role of hole and electron injection layers, respectively. Because of the excellent reflective properties of the silver films (that is, the reflectivity of the silver film with the thickness of 200 nm is more than 99% and that of 35 nm is about 50%), the parallel silver electrodes also double as the high-quality reflectors to form an optical Fabry-Pérot planar microcavity. Thin 2D-OSCs of 6M-DSB were chosen as the optical medium inside the microcavity, because of its giant anisotropy of the refractive index along Y (n = 2.40) and X (n = 1.95) directions (see Figure 1a and also text below)[42,43]. Furthermore, its semiconducting feature and high solid-state photoluminescence (PL) quantum yield (PLQY = 0.9931, see details in Supplementary Fig. S1) assure efficient electron and hole injection for high-performance EL.

Thin 2D-OSCs of 6M-DSB were prepared by the physical vapor transport method, with the lateral size in a millimeter scale and a thickness of about 990 nm (Supplementary Fig. S2). The uniform and smooth surface of 6M-DSB 2D-OSCs with a roughness less than 1.5 nm pave the way for fabricating microcavity CP-OLEDs. As-prepared 2D-thin-OSCs of 6M-DSB were then transferred on the 200-nm Ag film pre-deposited on silicon wafer. Finally, CsF/Ag films were vacuum deposited sequentially through a mesh-mask, giving rise to an array of microcavity CP-OLEDs on the surface of thin 2D-OSCs. Fig. 1a also presents the top-view bright-field photograph of patterned microcavity CP-OLEDs. It can be seen that subunits of CP-OLED array are in 40×40 μm$^2$ and separated by 20 μm to avoid cross-talk and short circuit. The thin 2D-OSCs of 6M-DSB adopt a lamellar structure with the crystal (001) plane parallel to the substrate (Fig. 1b and Supplementary Fig. S3)[44]. Within (001) plane, nearly planar 6M-DSB molecules stack in a brickwork arrangement among their short-axis with the nearest π-π distance ca. 3.6 Å. Their molecule long-axis is tilted at an angle of 63° respect to the substrate[44]. It can be seen from Fig. 1b that this brickwork arrangement brings about significant anisotropic molecular packing arrangements (molecular packing density) along and parallel to the π-π interaction (defined as Y- and X-direction, respectively), thus leading to a strong anisotropy of the refractive index. We associate the linear polarizations of the cavity modes along X- and Y-direction as X- and Y-polarization, respectively.

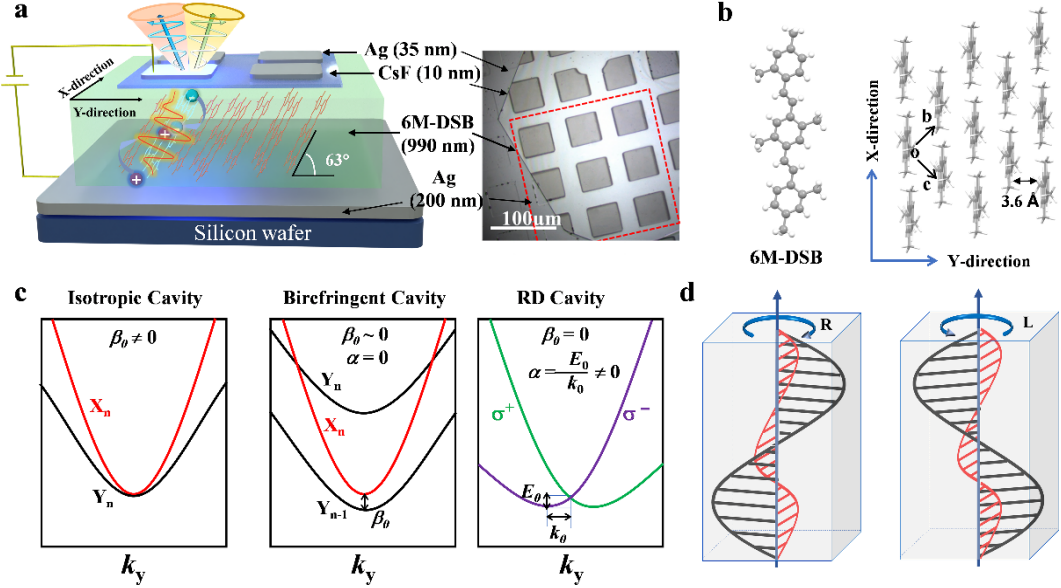

**Figure 1.** (a) Schematic diagram of the microcavity CP-OLED device structure. The bottom Ag (200 nm) and the top Ag (35 nm)/CsF (10 nm) films serve as the hole and electron injection layers for the OLED and meanwhile function as two mirrors forming a Fabry-Pérot planar microcavity. The 6M-DSB OSCs serve as the emitting layer of the OLED and as birefringent optical medium necessary for the realization of SOI in the microcavity. Right panel: the top-view bright-field of patterned microcavity CP-OLEDs. (b) Left: Molecular structure of 6M-DSB from single-crystal data, showing the enhanced planarity. Right: brickwork molecular packing arrangement within the (001) crystal plane, viewed perpendicular to the microribbon top-facet. The nearest intermolecular distance is d ≈ 3.6 Å. (c) Left: two orthogonally linearly polarized modes with the same parity in an isotropic microcavity. Middle: the dispersion of two orthogonally linearly polarized modes in an anisotropic microcavity. Right: RD SOI emerges when two orthogonally linearly polarized modes with opposite parity are resonant. (d) The resonant X- and Y-polarized cavity modes of opposite polarity. The 6M-DSB crystal in the microcavity hence acts as a half-wave plate, and the intrinsic mode polarization of the light emitting side of the mirror turns into a circle, corresponding to left-handed and right-handed circular polarizations, respectively.

In theory, such a birefringent microcavity can be approximately described by an effective Hamiltonian $H(\boldsymbol{k}) = H_{\text{TETM}} + H_{\text{RD}} + H_{\text{XY}}$, where $H_{\text{TETM}}$ describes the intrinsic transverse-electric-transverse-magnetic (TE-TM) splitting of the cavity modes[45], $H_{\text{RD}} = -2\alpha\hat{\sigma}_z k_y$ is the RD Hamiltonian[46,33,34], giving rise to a spin-splitting along $k_y$ direction with the strength $\alpha$, and $H_{\text{XY}} = \beta_0\hat{\sigma}_x$ is the

Hamiltonian representing the XY splitting[45], i.e., the energy splitting ($\beta_0$ at $\mathbf{k} = 0$) of the perpendicularly linearly polarized modes (X- and Y-polarizations) with opposite parity (here, we define it as $\beta_0 = E_X - E_Y$, where $E_X$ and $E_Y$ are the ground state energies of X and Y modes of opposite parity). The above effective Hamiltonian in the circular polarization basis can be written in the form of a 2×2 matrix:

$$H(\mathbf{k}) = \begin{pmatrix} E_0 + \frac{\hbar^2}{2m}\mathbf{k}^2 - 2\alpha k_y & \beta_0 + \beta_1 \mathbf{k}^2 e^{2i\varphi} \\ \beta_0 + \beta_1 \mathbf{k}^2 e^{-2i\varphi} & E_0 + \frac{\hbar^2}{2m}\mathbf{k}^2 + 2\alpha k_y \end{pmatrix}, \quad (1)$$

where $E_0$ is the energy of the ground state, $m$ is the effective mass of cavity photons, $\beta_1$ is the strength of the TE-TM splitting, and $\varphi$ ($\varphi \in [0, 2\pi]$) is the polar angle.

The left panel of Fig. 1c depicts the energy ($E$) vs. momentum ($k$) dispersion in an isotropic microcavity. Here, two orthogonally polarized cavity modes with the same parity ($X_n$, $Y_n$) degenerate at $k = 0$, with $\beta_0 > 0$ ($\beta_0 = 0$ means two orthogonally polarized cavity modes with different parity at $k = 0$ to be resonant). Once 2D-OSCs of 6M-DSB were inserted, the anisotropy of its refractive index leads to a birefringent microcavity, in which the splitting between $X$- and $Y$-polarized modes at $k = 0$ occurs, leading to a reduction of $\beta_0$ (see the middle panel of Fig. 1c). By carefully tuning the cavity length, i.e., the thickness of the 2D-OSCs, two orthogonally linearly polarized cavity modes with opposite parity ($X_n$, $Y_{n-1}$) approach each other at $k = 0$, that is, $\beta_0 \sim 0$. In the resonant case ($\beta_0 = 0$), a clear spin-splitting appears along $k_y$ direction due to RD SOI with $\alpha \neq 0$ (see right panel of Fig. 1c) as predicted by Eq (1) and therefore linear polarizations change to circular polarizations. That is, at resonance of $X_n$ and $Y_{n-1}$, the phase difference between the $X$- and $Y$-polarized modes across the intracavity anisotropic 2D-OSC rotate by π, such that the anisotropic 2D-OSC act as a half-wave plate and result in a change of the eigenmode polarization at the mirror interfaces from linear to circular polarizations (Fig. 1d)[33].

To experimentally demonstrate the above theoretical prediction, we firstly measured the unpolarized angle-resolved reflectivity (ARR) of our microcavity CP-OLED along $Y$-direction. The reflectivity is plotted as a function of wavelength (or energy) and angle (or momentum $k_y$), measured by using a home-made micro-scale ARR measurement system (Supplementary Scheme S4). Fig. 2a present two sets of modes with distinctive curvatures. The black and red dashed lines are simulated dispersions, in good agreement with experimental results. The simulated refractive index of the two cavity modes (that is, 1.95 for $X$-polarized modes and 2.40 for $Y$-polarized modes) also support the fact of the giant anisotropy of the thin 2D-OSCs of 6M-DSB (Supplementary Fig. S4). We measure the polarized reflection spectrum and obtain accurate refractive index values (Supplementary Fig. S10), which are in line with our simulation results. We also perform angle-resolved PL (ARPL) measurements under the excitation of a 405-nm continuous-wave laser. The ARPL (Fig. 2b) and ARR (Fig. 2a) spectra show exactly the same dispersions. Furthermore, polarization-resolved ARPL experiments have been carried out by adding a quarter-wave plate and a linear polarizer along the detection optical path

(Supplementary Scheme S5), which allows us to distinguish the polarization of the relevant optical modes. The modes with larger (red dashed lines) and smaller (black dashed lines) curvatures are *X*- and *Y*-polarization corresponding to *X*- and *Y*-direction, respectively, of the 2D-OSCs of 6M-DSB, which are both well consistent with the calculated results of the cavity modes by using the 2D cavity photon dispersion relations[34].

The energy splitting values $β_0$ between the X- and Y-polarized modes present wavelength-dependent characteristics, for example, $β_0$ is about 23 meV, 0 meV, and -21 meV for near 448 nm, 497 nm, and 558 nm, respectively, as shown in Fig. 2a and b. Different from other successive dispersion branches (such as at 448 nm and 558 nm), the crossing point appears at 497 nm and $k_y = 0$ originated from $X_9$ and $Y_8$ branches (black dotted circle in Fig. 2a), which suggests that the RD SOI emerges in this device. To investigate the polarization property of these cavity modes, we further measured the Stokes vector components[47] to analyze the pseudospin behaviors. The $S_1$ components of the Stokes vector of $X_8$ and $Y_7$ branches show strongly linearly polarized PL emission (Fig. 2e), while their corresponding $S_3$ components exhibit relatively weak (Fig. 2f). As *X*- and *Y*-polarized modes approach, the near resonant $X_9$ and $Y_8$ branches at $k_y = 0$ enhances the RD SOI, leading to the clear splitting of the paraboloids in $k_y$ direction (Fig. 2b). The $S_3$ components present two separate circles with opposite signs and they are highly circularly polarized with much higher polarization degree (Fig. 2d), while the $S_1$ components become very weak (Fig. 2c). Therefore, the PL-active CP emission due to the RD SOI has been clearly demonstrated in our microcavity OLEDs.

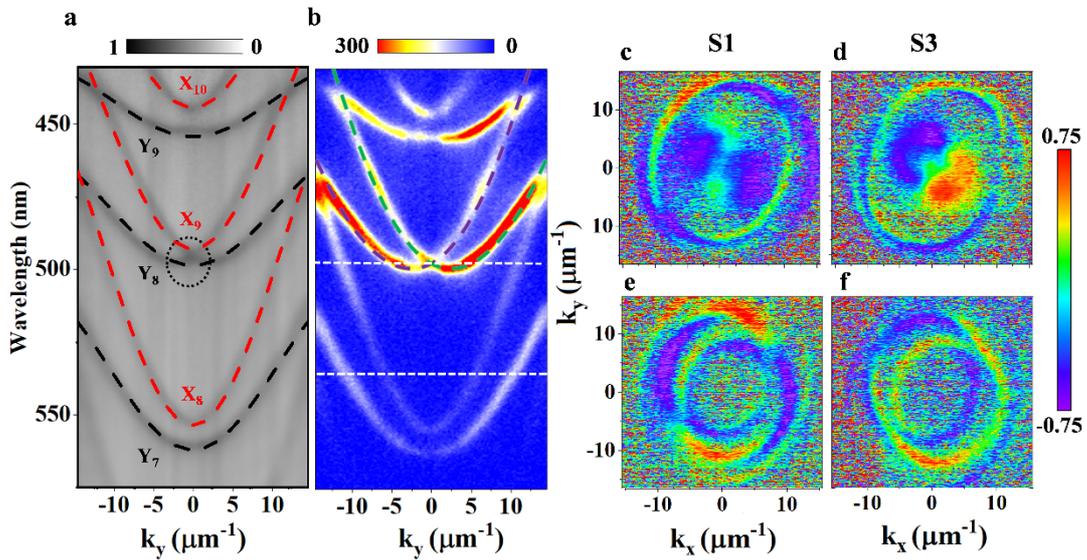

**Figure 2.** (a) ARR and (b) ARPL of the microcavity with the organic-layer thickness of 990 nm. The dotted curves in (a) represent the simulated X and Y cavity modes, respectively. The dashed curves in (b), indicating the split CP modes, are calculated by solving Eq. (1) with the parameters $E_0 = 2.495$ eV (497 nm), $m = 2.2 \times 10^{-5} m_e$ ($m_e$ is the free electron mass), $β_0 = 0$, $β_1 = 0.7$ meV, $α = 37$ eV·Å. (c-f) Cross-section maps of the 2D tomography at 497 nm and 535 nm in momentum

space, corresponding to the dashed lines in (b), respectively. $S_1$ components of the Stokes vector at 497 nm (c) and 535 nm (e), $S_3$ components of the Stokes vector at 497 nm (d) and 535 nm (f).

The EL performances of our microcavity CP-OLED were then investigated. Fig. 3a shows the energy level diagram of the OLED architecture. Ag and Ag/CsF are used as the upper and lower electrodes to inject holes and electrons, respectively. Under a certain voltage, a device achieves uniform and bright EL and edge waveguide (Fig. 3b and Supplementary Figure S5). We collect the EL emissions from the upper electrode and the crystal edge and compare them with the PL spectrum of 6M-DSB OSCs (Fig. 3c). The waveguided EL light emitted from the OLED is trapped in single crystals in the form of a whisper gallery mode and then propagates out of the crystal edge. So, this EL spectrum (middle panel of Fig. 3c) is coincident with the PL spectrum of the OSCs (upper panel of Fig. 3c). Notably, the EL spectrum from the upper electrode (bottom panel in Fig. 3c) exhibits multiple microcavity resonant peaks, and these peaks' wavelengths are obviously consistent with the PL spectrum of the OSCs. This indicates that the EL emission comes from the OSC active layer of the OLED and is regulated by the microcavity. The EL performances of a single-crystal OLED based on 6M-DSB single crystals are shown in Fig. 3d and 3e. The maximum luminance and current efficiency of about 60000 cd/m$^2$ and 1.48 cd/A, respectively, were obtained. Thanks to the good crystal quality of the OSCs, our OLED can withstand high voltage of tens of volts (that is, current density of 22.7 A/cm$^2$) and stably emit light without any damage. The highest luminance of our OLED reaches 60000 cd/m$^2$ at the current density of 7.6 A/cm$^2$, which is one of the highest luminance of single-crystal OLEDs.

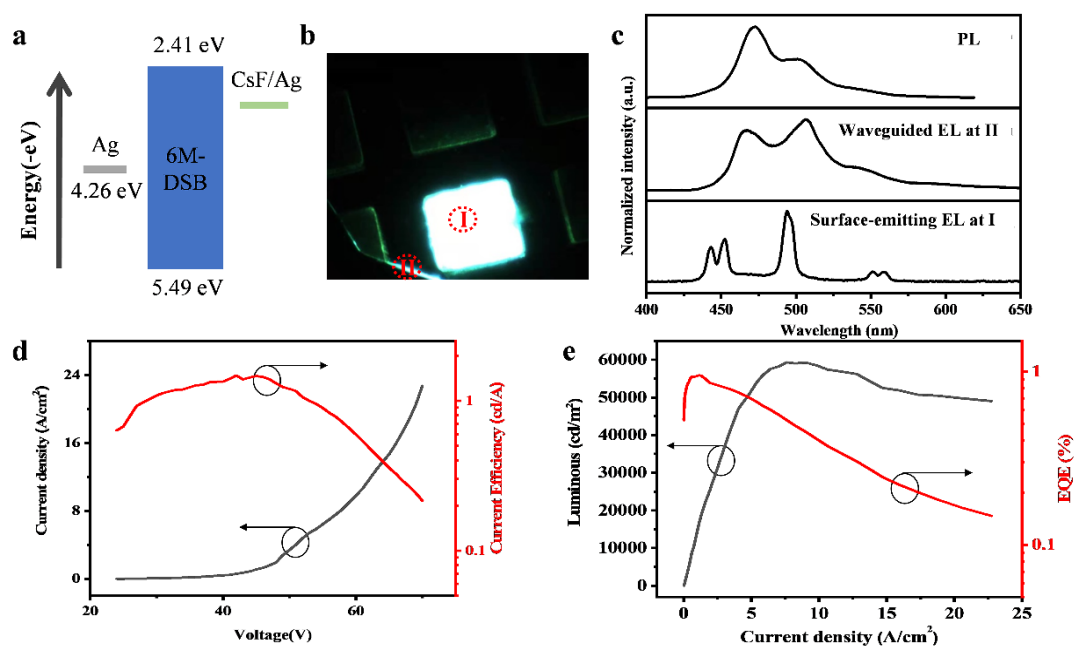

**Figure 3.** (a) Energy level diagram of the device. (b) Photomicrograph of the 6M-DSB single crystal OLED array under EL. (c) Upper: PL spectrum of the

6M-DSB crystal. Middle: Spectrum of the crystal edge waveguide under EL as shown in (b). Lower: Spectrum of light emanating from the microcavity under EL as shown in (b). (d) Dependence of the current density (black line) and the current efficiency (red line) on the voltage. (e) Dependence of the luminescence (black line) and the EQE (red line) on the current density.

In order to study the influence of the RD SOI on the EL emission, we perform the angle-resolved EL (AREL) measurement for our OLED. The angle range of the AREL spectrum is only ±15° limited by numerical aperture of the microscope lens in our setup. The obtained AREL spectrum also shows a clear RD spin-splitting, which matches well with the EL spectrum (Supplementary Figure S6). We further performed the circularly polarized (σ+ and σ−) AREL spectroscopy. Expectedly, the EL emission exhibit strong left- and right-handed circular polarization in the vicinity of 497 nm at $k = 0$ as shown in Fig. 4a and 4b, respectively, which agrees to the result of the ARPL as shown in Fig. 4c and 4d. This strongly testifies that the RD SOI occurs at the condition of electrical excitation and induces the active CP emission due to the spin splitting. Correspondingly, the EL emission spectra of unpolarized, σ+ and σ− are shown in Fig. 4e. The left- and right-handed circular polarized emissions are exactly coincident with the above analysis of the cavity modes.

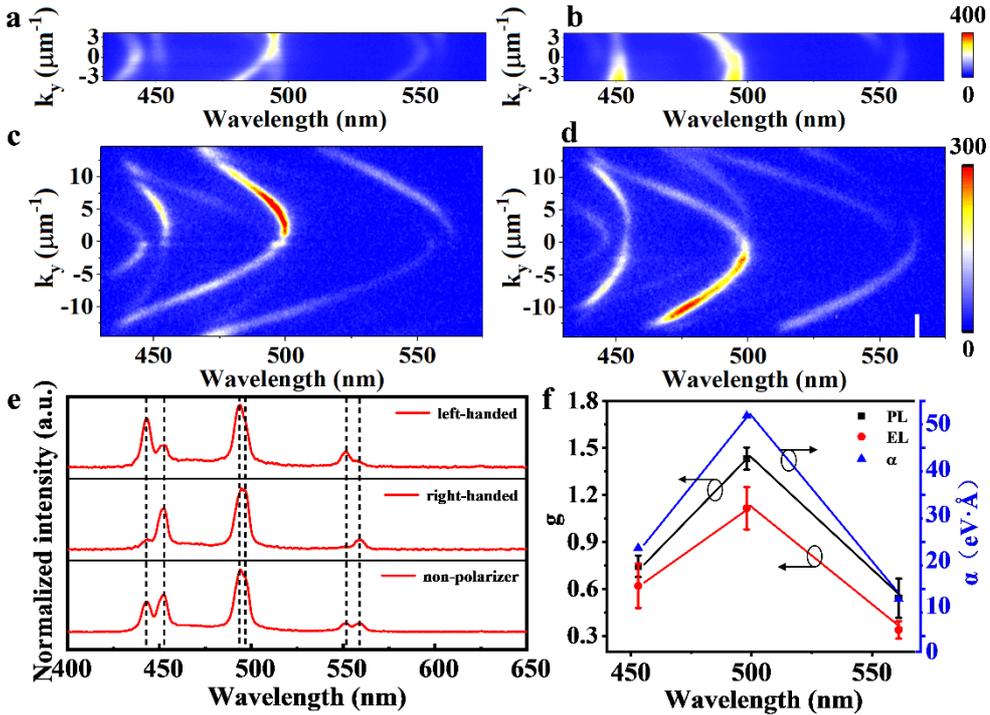

**Figure 4.** Polarized angle-resolved spectra of EL (a,b) and PL (c,d). (e) EL emission spectra of unpolarized, σ+ and σ− components. (f) Wavelength-dependent spin-splitting coefficient α of the RD effect and circularly polarized luminescence dissymmetry factor g of PL and EL.

We calculate the RD spin-splitting parameters ($\alpha = 2E_0/k_0$)[48], as a typical value for the feature of RD SOI, at different photonic modes and show in Fig. 4f (blue triangles). Two modes with different polarizations and opposite parity are brought into resonance and give rise to wavelength-dependent α values of the RD effect (Supplementary Fig. S5). In the vicinity of the resonance ($X_9$ and $Y_8$ modes), we obtain a giant RD parameter of α = 52 eV·Å, which is much larger than that reported in liquid-crystal optical cavities[33]. The key parameters of circular polarization asymmetry factors ($g_{EL}$ for EL and $g_{lum}$ for PL) are usually employed to characterize the performance of CP-OLEDs and can be defined as $g_{EL}$ (or $g_{lum}$) = 2 × ($I_L − I_R$)/($I_L + I_R$), where $I_L$ and $I_R$ correspond to the intensities of left- and right-handed polarization, respectively[49]. The obtained $g_{EL}$ and $g_{lum}$ on the different photonic modes are presented in Fig. 4f. At the resonance wavelength, $g_{EL}$ and $g_{lum}$ reach the maximum values of about 1.4 and 1.1, respectively, which are both the best performances of CP-OLEDs. Our device is also the first single-crystal CP-OLED with highly luminescent CPEL based on organic small molecules.

**Conclusions**

In summary, we propose a new strategy for designing chiral-emitter-free single-crystal CP-OLEDs with high luminance and large $g_{EL}$ value through introducing artificial RD SOI into optical microcavities. Thanks to high emission efficiency and high carrier mobility of the 6M-DSB OSC, our microcavity CP-OLEDs exhibit a maximum luminance of exceeding 60000 cd/m$^2$ and a large $g_{EL}$ value of 1.1, which are among the best performances of single-crystal CP-OLEDs. The introduction of organic single crystal to such EL devices brings about two advantages: i) the highly ordered molecular stacking arrangement in single crystals can result in anisotropic refractive index in different directions, which benefit the occurrence of the RD effect. ii) The high current density in single-crystal OLEDs (e.g. 69.25 A cm$^{-2}$ in our case, which is three orders of magnitude higher than that of thin-film OLEDs) also makes them potential candidates also for the future realization of organic electrically-pumped solid-state lasers. Our new strategy provides a promising design for efficient single-crystal CP-OLEDs, which may greatly promote the development of CP-OLEDs with use in future 3D display applications.

**Acknowledgements**

This work was supported by the National Key R&D Program of China (Grant No. 2018YFA0704805, 2018YFA0704802 and 2017YFA0204503), the National Natural Science Foundation of China (22150005, 22090022, 21833005 and 21873065), the Natural Science Foundation of Beijing, China (KZ202110028043), Beijing Talents Project (2019A23), Capacity Building for Sci-Tech Innovation-Fundamental Scientific Research Funds, Beijing Advanced Innovation Center for Imaging Theory and Technology. The Paderborn group acknowledges support by the Deutsche

Forschungsgemeinschaft (DFG) through the collaborative research center TRR142 (project A04, No. 231447078).

The authors thank Dr. H. W. Yin from ideaoptics Inc. for the support on the angle-resolved spectroscopy measurements.

**Author contributions**

J.-C.J.,C.X., J.-B.D. and Q.L. designed the experiments and performed experimental measurements. X.M. and S.S. performed the theoretical calculation and analysis. X.M., Q.L. and H.-B.F. wrote the manuscript with contributions from all authors. Q.L. and H.-B.F. supervised the project. All authors analyzed the data and discussed the results.

**Competing interests**

The authors declare no competing interests.

**Additional information**

Correspondence should be addressed to Q.L. and H.-B.F.: liaoqing@cnu.edu.cn, hbfu@cnu.edu.cn

# Supplementary Materials

**MATERIALS AND METHODS**

**1. Synthesis of 6M-DSB**

The compound used in our work, 1,4-bis((E)-2,4-dimethylstyryl)-2,5-dimethylbenzene (6M-DSB), was synthesized (Scheme S1) according to Horner-Wadsworth-Emmons reaction (F. Gao *et al.*, *Angew. Chem. Int. Ed.* 2010, 49, 732-735. & Z. Xu *et al.*, Adv. Mater. 2012, 24, OP216-OP220.). All starting materials were purchased from Sigma-Aldrich and used as received without further purification. The tetrahydrofuran (THF, HPLC grade) and hexane were purchased from Beijing Chemical Agent Ltd., China. Ultra-pure water with a resistance of 18.2 MΩ·cm$^{-1}$ were used in all experiments, produced by Milli-Q apparatus (Millipore).

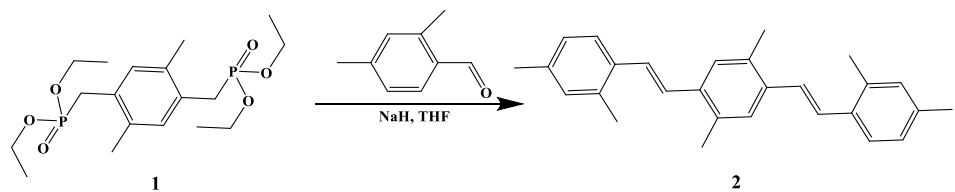

**Scheme S1.** The synthesis of the 6M-DSB molecule.

A mixture of 2,5-dimethyl-1,4-xylene-bis(diethyl phosphonate) (1.06 g, 2.61 mmol) and the 2,4-dimethylbenzaldehyde (5.74 mmol) in tetrahydrofuran (THF) cooled in an ice bath was added 2 eq. NaH in small portions during a 30 min period. The reaction mixture was stirred at room temperature for 3 hours and poured into water. The phase was extracted with CH$_2$Cl$_2$. The pooled organic phases were washed with water, dried over anhydrous MgSO$_4$, filtered, and evaporated. The product was separated by flash chromatography on silica gel by means of CH$_2$Cl$_2$/petroleum ether (1:4). Finally a highly fluorescent powder was obtained as the title compound (573 mg) in 85% yield. The obtained 6M-DSB molecule was characterized by $^1$H NMR (Scheme S2).

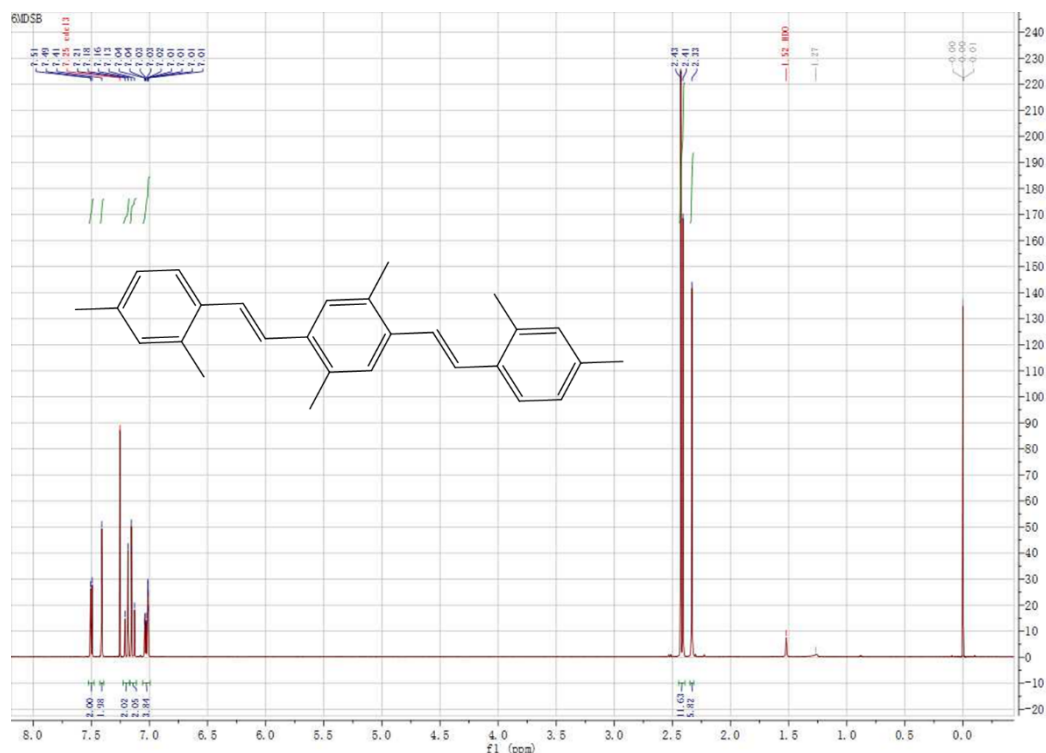

**Scheme S2.** ¹H Nuclear magnetic resonance（NMR）spectrum of 6M-DSB.
1H NMR (600 MHz, Chloroform-d) δ 7.50 (d, J = 7.8 Hz, 2H), 7.41 (s, 2H), 7.20 (d, J = 15.9 Hz, 2H), 7.14 (d, J = 16 Hz, 2H), 7.06 – 6.99 (m, 4H), 2.42 (d, J = 11.1 Hz, 12H), 2.33 (s, 6H).

## 2. The preparation of 6M-DSB single crystals

In our experiment, 6M-DSB single crystals were fabricated using a facile physical vapor deposition (PVD) method. A quartz boat carrying 3 mg 6M-DSB was then placed in the center of a quartz tube which was inserted into a horizontal tube furnace. A continuous flow of cooling water inside the cover caps was used to achieve a temperature gradient over the entire length of the tube. To prevent oxidation of 6M-DSB, Ar was used as inert gas during the PVD process (flowrate: 100 sccm·min$^{-1}$). The pre-prepared hydrophobic substrates were placed on the downstream side of the argon flow for product collection and the furnace was heated to the sublimation temperature of 6M-DSB (at temperature region of ~ 230 °C), upon which it was physically deposited onto the pre-prepared hydrophobic substrates for 3 hours.

The thickness of 6M-DSB crystal can be controlled by the PVD method in our experiments. For example, as presented in Scheme S3, the crystal thickness obtained is about 380 nm when the deposition temperature is set to 150 °C and the flow rate of inert gas is set to 100 sccm·min$^{-1}$. As the deposition temperature is raised to 110 °C, the crystals with thickness of about 850 nm are obtained. When deposition temperature reaches 90 °C, the crystal thickness increases to about 1300 nm.

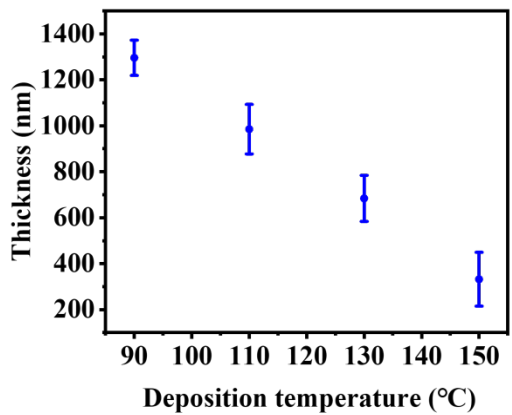

**Scheme S3.** Dependence of the crystal thickness on the deposition temperature.

### 3. The preparation of 6M-DSB single-crystal CP-OLEDs

Firstly, we use the metal vacuum deposition system (Amostrom Engineering 03493) to thermally evaporate a silver film with the thickness of 200 ± 5 nm (reflectivity R ≥ 99%) on a silicon wafer substrate, 6M-DSB crystals prepared by PVT are transferred to a silver substrate using a mechanical transfer method. The 6M-DSB crystals were uniformly dispersed on the silver film substrate. On this basis, the upper array of 10 nm cesium fluoride and 35 nm silver (R ≈ 50%) was prepared by the method of copper mesh mask to form microcavities. The 10 nm cesium fluoride layer is used to reduce the injection barrier between the electrode and the semiconductor layer to achieve good electron injection.

### 4. Structural and spectroscopic characterization

As-prepared 6M-DSB crystals were characterized by transmission electron microscopy (TEM, JEOL, JEM-2100) in which PVD grown crystals were mechanically transferred to a carbon-coated copper grid for testing. TEM measurement was performed at room temperature at an accelerating voltage of 100 kV. The X-ray diffraction (XRD, Japan Rigaku D/max-2500 rotation anode X-ray diffractometer, graphite monochromatized Cu $K_\alpha$ radiation ($\lambda$ =1.5418 Å)) operated in the 2θ range from 3° to 30°, by using the samples on a cleaned glass slide.

The fluorescence micrograph, diffused reflection absorption and emission spectra were measured on Olympus IX71, HITACHI U-3900H, and HITACHI F-4600 spectrophotometers, respectively. The photoluminescence spectrum of the device was characterized by using a homemade optical microscope equipped with a 50 × 0.9 NA objective (Scheme S4). A single selected device is excited on a two-dimensional (2D) movable table using a continuous laser focused at 405 nm to a 50-μm diameter spot. Spatially resolved PL spectra were collected underneath by using a 3D-movable objective and detected using a liquid-nitrogen cooled charge-coupled device (CCD).

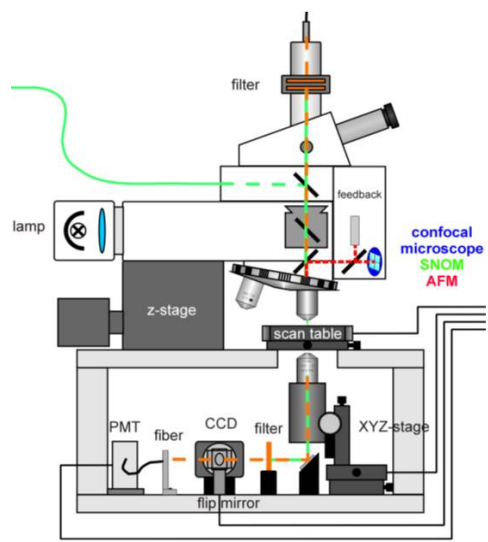

**Scheme S4.** Schematic demonstration of the experimental setup for the optical characterization: the near-field scanning optical microscopy.

## 5. The angle-resolved spectroscopy characterization

The angle-resolved spectroscopy was performed at room temperature by the Fourier imaging using a 100× objective lens of a NA 0.95, corresponding to a range of collection angle of ±60° (Scheme S5). The incident white light of the Halogen lamp with the wavelength of 400-700 nm is used to focus on the region to be measured. The k-space or angular distribution of the reflected light was located at the back focal plane of the objective lens. Lenses L1-L4 formed a confocal imaging system together with the objective lens, by which the k-space light distribution was first imaged at the right focal plane of L2 through the lens group of L1 and L2, and then further imaged, through the lens group of L3 and L4, at the right focal plane of L4 on the entrance slit of a spectrometer equipped with a liquid-nitrogen-cooled CCD. The use of four lenses here provided flexibility for adjusting the magnification of the final image and efficient light collection. Tomography by scanning the image (laterally shifting L4) across the slit enabled obtaining spectrally resolved two-dimensional (2D) k-space images.

In order to investigate the polarization properties, we placed a linear polarizer, a half-wave plate and a quarter-wave plate in front of spectrometer to obtain the polarization state of each pixel of the k-space images in the horizontal-vertical (0° and 90°) and circular (σ+ and σ− ) basis (S. Dufferwiel *et al.*, *Phys. Rev. Lett.* 2015, 115, 246401. & F. Manni *et al.*, *Nat. Commun.* 2013, 4, 2590.).

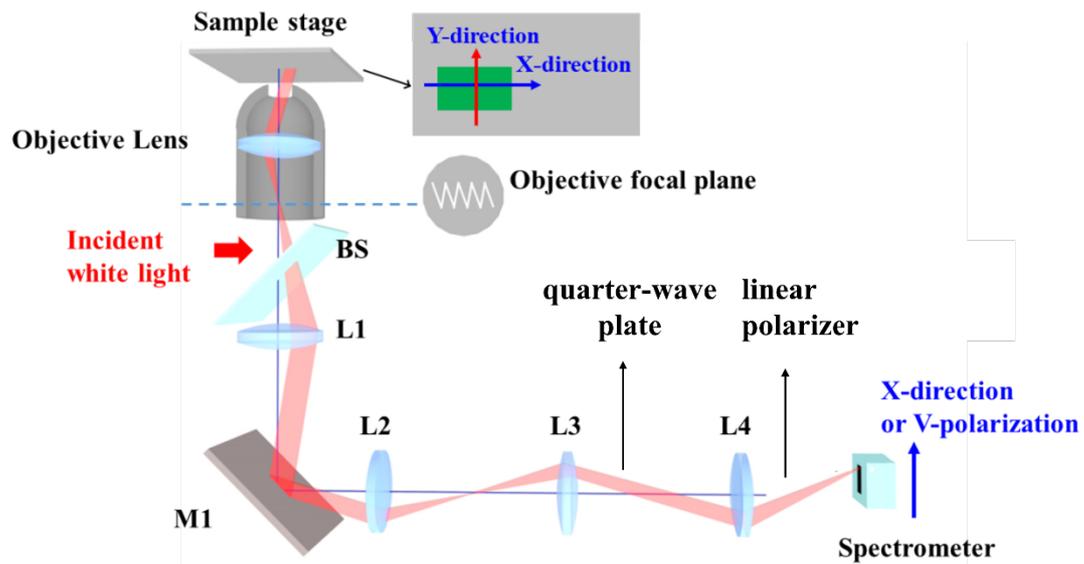

**Scheme S5.** Experimental setup for the polarization-resolved PL and EL spectra. BS: beam splitter; L1-L4: lenses; M1: mirror. The red beam traces the optical path of the reflected light from the sample at a given angle.

**PLQY of 6M-DSB single crystals**

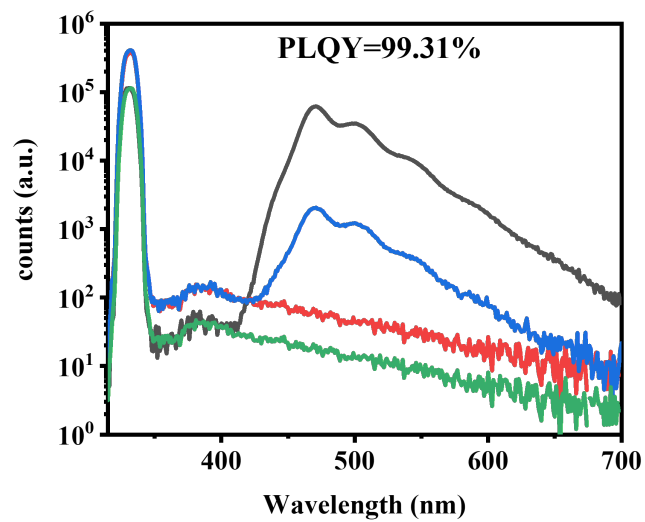

**Figure S1.** PLQY of 6M-DSB single crystals was measured through an absolute method by using an integration sphere in FLS-1000. At the excitation wavelength of 330 nm, PLQY of 6M-DSB crystals is determined to be 0.9931.

**Morphology of 6M-DSB single crystals grown by the PVT method**

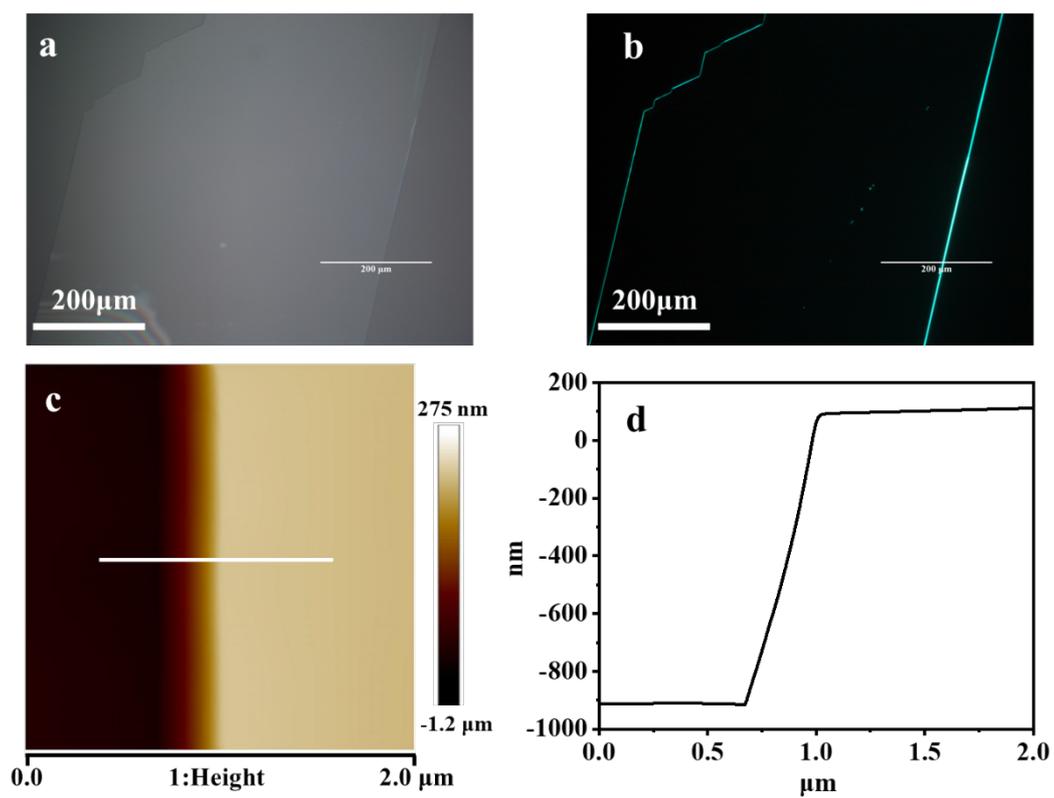

**Figure S2.** Microscopic bright field and fluorescence images of the 6M-DSB crystal prepared by physical vapor deposition method are shown in (a) and (b). (c) The AFM image of the crystal used for device preparation. (d) The morphology curve along the white line in (c).

**Characteristics of crystalline structure of 6M-DSB single crystals**

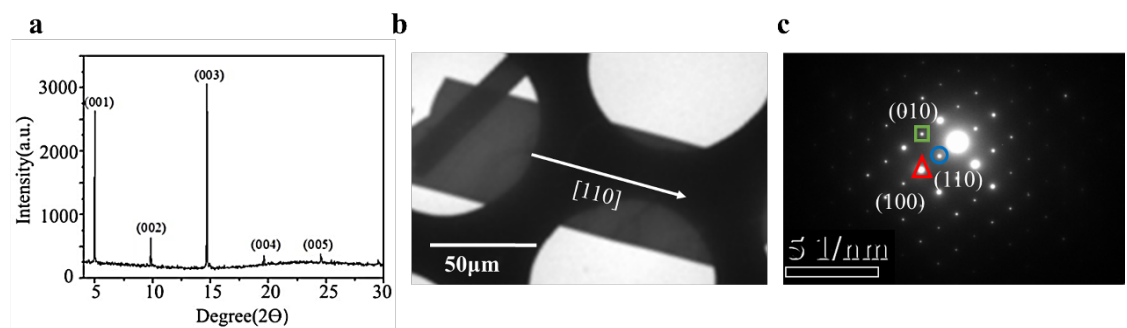

**Figure S3.** (a) XRD pattern of ensemble 6M-DSB microribbons filtered on the surface of an alumina membrane. (b) TEM image of a typical 6M-DSB microribbon. (c) SAED pattern of the microribbon in (b).

The XRD pattern was measured by a D/max 2400 X-ray diffractometer with Cu Kα radiation ($\lambda$ = 1.54050 Å) operated in the 2θ range from 4° to 30°. According to the single-crystal data, the monoclinic crystal of 6M-DSB has the lattice parameters of $a$ = 4.7533(10) Å, $b$ = 5.9928(12) Å, $c$ = 18.235(4) Å, α = 96.08(3)°, β = 96.46(3)°, and γ = 90.15(3)°. The XRD spectrum of microribbons shows a series of peaks corresponding to the (001) crystal plane with an interplanar spacing of 18.15 Å (Figure S3a). The observation of high-order diffraction peaks, such as (002)-(005), suggests that the crystal adopts a lamellar structure with the crystal (001) plane being parallel to the substrate. Figure S3c presents SEAD pattern recorded by directing the electron beam perpendicular to the flat surface of a single microribbon. The clearly observed SAED spots and its rectangular symmetry suggest that 6M-DSB microribbons are single crystals. The squared and triangled sets of SAED spots correspond to (020) and (100) crystal planes with $d$-spacing values of 6.10 and 4.76 Å, respectively, and the blue circled set of SAED spots are attributed to the (110) crystal plane with a $d$-spacing value of 3.67 Å. Combining the XRD, SAED and TEM results (Figure 3b) together, the 6M-DSB ribbons grow along the [110] crystal direction, bound by (001) and (0-10) crystal planes on the top and bottom surfaces and (1-10) and (-110) crystal planes on the lateral surfaces.

**Simulation of two cavity modes**

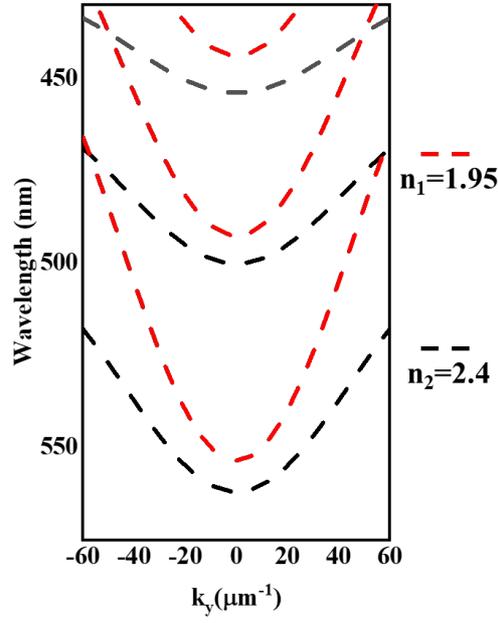

**Figure S4.** The corresponding refractive index of the two simulated cavity modes (i.e., *Y*-polarized mode and *X*-polarized mode).

We have calculated X- and Y-polarized modes by using the two-dimensional cavity photon dispersion relations (ref. J. Ren, *et al.*, *Laser Photon. Rev.* 2022, 16, 2100252.). According to the equation,

$$E_{CMn}(\theta) = \sqrt{\left(E_c^2 \times \left(1 - \frac{\sin^2\theta}{n_{eff}^2}\right)^{-1}\right) - (n-1) \times l}$$

Where $\theta$ represents the incidence angle, $E_{CMn}(\theta)$ is the cavity photon energy of the $n^{th}$ cavity mode as a function of $\theta$, $E_c$ represents the cavity modes energy at $\theta = 0°$, $E_{CM1}(\theta)$ represents the energy of the first cavity mode when $n = 1$, $(n-1) \times l$ represents the energy difference from the first cavity mode. The refractive indices are calculated to be 1.95 and 2.40 for the red and black curves, respectively.

## OLED array of 6M-DSB crystals

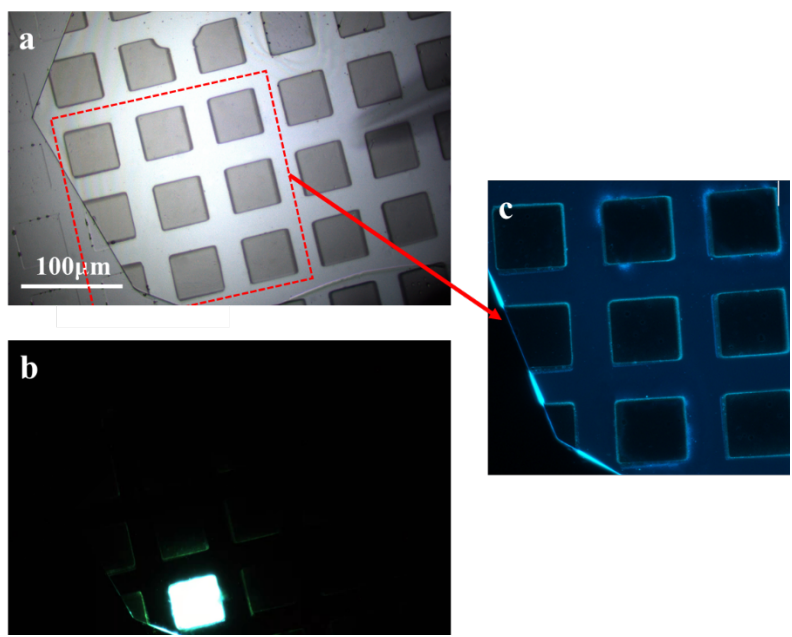

**Figure S5.** (a) Microscopic bright field image of 6M-DSB crystal OLED array. (b) Electroluminescent photographs of individual devices. Under electroluminescence, the device exhibits uniformly bright electroluminescence and edge waveguide. (c) Microfluorescence image of the red box area in Figure a.

**Angular resolved spectra of the device**

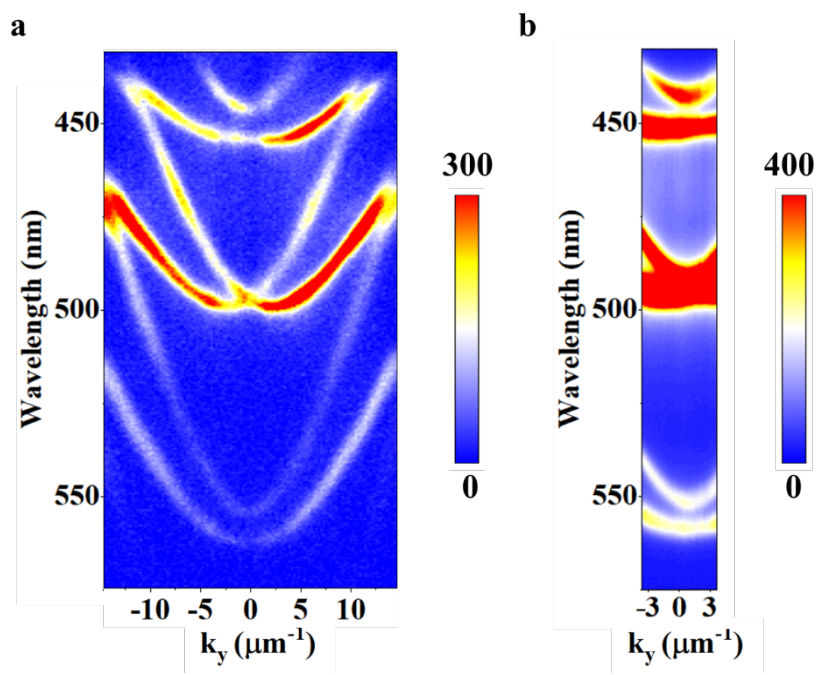

**Figure S6.** (a) Angle-resolved PL spectra under 405-nm laser excitation. (b) Angle-resolved EL spectra of the same device.

**Angular resolved spectra of the devices with the organic-crystal thickness of 835 nm, 990nm and 1325 nm.**

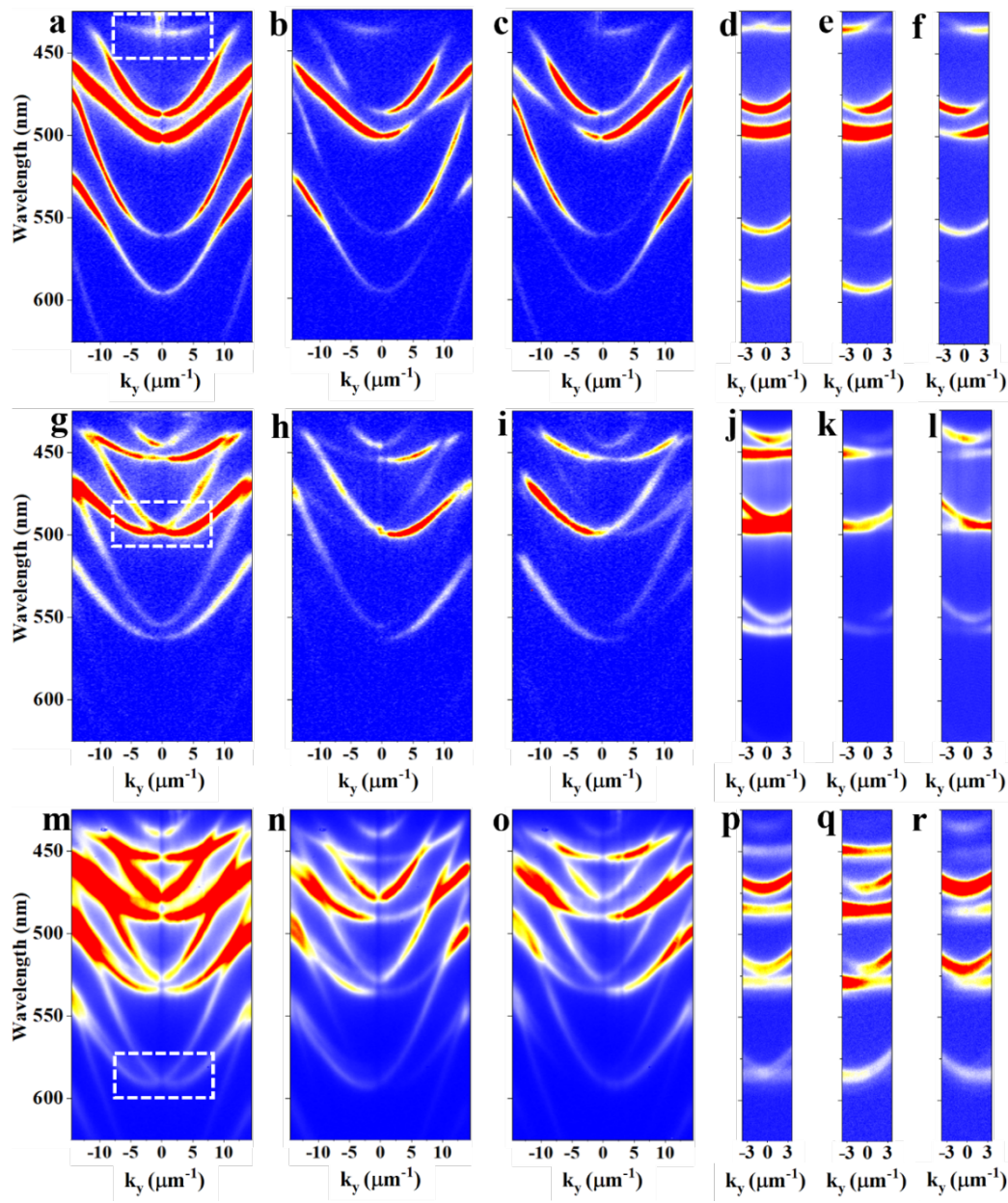

**Figure S7.** Angle-resolved photoluminescence (a,g,m) and electroluminescence spectra (d,j,p) of the organic-crystal microcavity with the crystal thickness of 835 nm, 990 nm and 1325 nm, respectively. The white dashed rectangles mark the positions where the RD effect occurs. The corresponding circularly polarized angle-resolved spectra of PL (b,c) and EL (e,f) with the organic-crystal thickness of 835 nm, PL (h,i) and EL (k,l) with the organic-crystal thickness of 990 nm, and PL (n,o) and EL (q,r) with the organic-crystal thickness of 1325 nm.

**Table S1. Summary of PL and EL Performance of the Devices with Different Thickness**.

| Thickness (nm) | $V_{on}$ (V) | Current density (A/m$^2$) | Luminous (cd/m$^2$) | EQE (%) | $g_{EL}$ | $g_{PL}$ |
|---|---|---|---|---|---|---|
| 835 | 19 | 36.77 | 60829 | 0.84 | -1.20≤ $g_{EL}$ ≤1.14 | -1.27≤ $g_{PL}$ ≤1.42 |
| 990 | 20 | 22.7 | 59324 | 0.96 | -1.25≤ $g_{EL}$ ≤1.11 | -1.30≤ $g_{PL}$ ≤1.43 |
| 1325 | 31 | 10.01 | 53868 | 1.008 | -1.23≤ $g_{EL}$ ≤1.07 | -1.31≤ $g_{PL}$ ≤1.40 |

In order to investigate the relation between the thickness of 6M-DSB crystal and the dissymmetry factor for circularly polarized emission, we chose three typical 6M-DSB crystals with the thickness of 835 nm, 990 nm and 1325 nm and performed their angle-resolved photoluminescence and electroluminescence spectra. As shown in Figure S7, the location of the crossing point, characteristic feature of the RD effect, shifts from shorter wavelength to longer wavelength as the increase of the crystal thickness. We calculated the dissymmetry factors of these three devices and the results are summarized in Table S1. The photoinduced and electroinduced dissymmetry factors are almost the same as those of 990-nm devices in the manuscript. Therefore, it can be concluded that the crystal thickness affects only the position of the crossing point induced by the RD spin-splitting not the dissymmetry factor.

**Reflection Spectrum and Refractive Index of 6M-DSB Single Crystal**

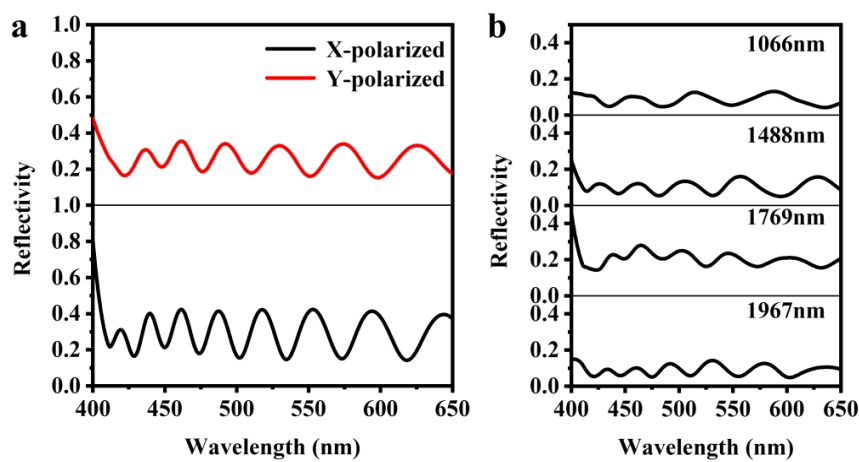

**Figure S8**. (a) Reflection spectra of the 2126-nm-microbelt cavity at X-polarization (black line) and Y-polarization (red line) (b) X-polarized reflection spectra of organic cavities with thickness of 1066 nm, 1488 nm, 1769 nm and 1967 nm.

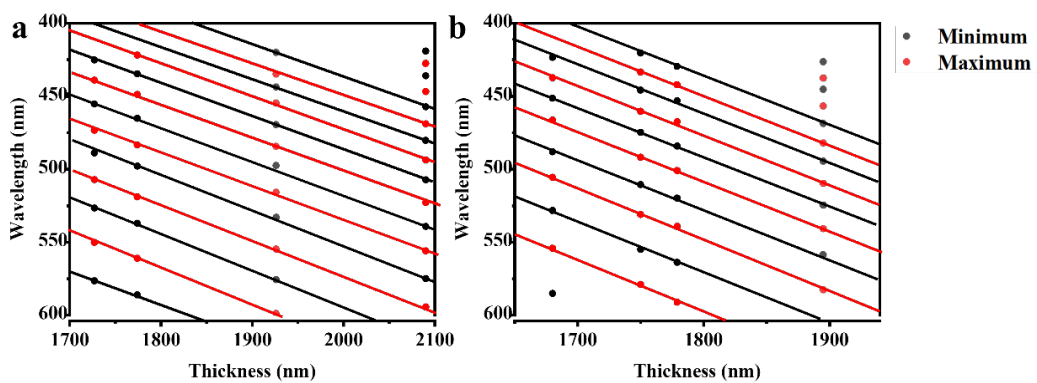

**Figure S9**. Wavelength of the interference maximum (red points) and minimum (black points) observed in reflection spectra of X-polarization (a) and Y-polarization (b).

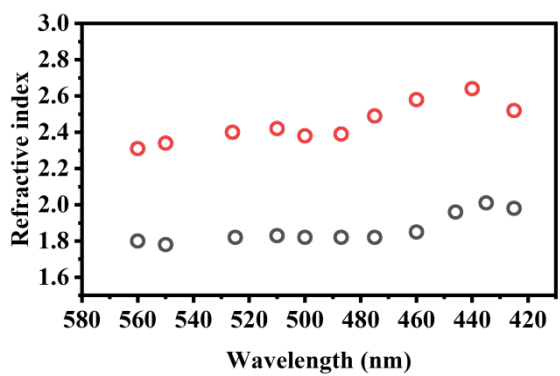

**Figure S10**. The calculated refractive index of 6MDSB microbelts in X-direction (black dot) and Y-direction (red dot).

We measured the refractive index using the method in the reference[1]. Firstly, the 6M-DSB microbelts are placed on the quartz substrate, and the Fabry-Pérot interference is formed on the top and bottom smooth surfaces. The reflection spectra of the samples are measured with polarized light on parallel to and perpendicular to the long direction of the crystal. Figure S8 shows the reflection spectrum of the sample with the thickness of 2126 nm, where the red line and black line represent the reflection spectra parallel to and perpendicular to the long direction of the crystal, respectively. The interference peak spacing in X- and Y-direction does not change significantly.

Figure S8b shows X-polarized reflection spectra of organic cavities with thickness of 1066 nm, 1488 nm, 1769 nm and 1967 nm, respectively. The interference conditions are given by $2n(\lambda)d = m\lambda$, where $n(\lambda)$ is the refractive index at wavelength $\lambda$, $m$ is the order of interference, and $d$ is the crystal thickness. In the reflection spectra, the interference minimum occurs when $m$ is an integer and the maximum occurs when m is a half integer. The wavelengths of the interference maximum and minimum were extracted from the reflection spectra (Figure S8) measured when the polarization of incident light parallel to and vertical to the long direction of microbelts and are plotted as a function of crystal thickness (Figure S9a). The black points correspond to the minimum and the red points correspond to the maximum. The black (red) lines are fitted by the black (red) points. Since two adjacent black (or red) lines have difference 1 in the interference order, the order $m$ can be determined by $m = d_1/(d_2-d_1)$, where $d_1$ and $d_2$ ($d_2 > d_1$) are the thicknesses corresponding to the order $m$ and $m+1$, respectively, at a fixed wavelength $\lambda$. The calculated $n(\lambda)$ is shown in Figure S10. The $n(\lambda)$ in the direction parallel to the long direction of the microbelt rises from 2.31 at 560 nm to 2.52 at 425 nm, and rises from 1.80 at 560 nm to 1.98 at 425 nm in vertical direction (Figure S10). This is basically consistent with our simulation results.